\documentclass[conference]{IEEETran}
\usepackage{graphicx}
\usepackage{cite}
\usepackage{amsmath,amssymb}
\usepackage{caption}
\usepackage{subcaption}
\usepackage{xcolor}
\usepackage[braket, qm]{qcircuit}
\usepackage{hyperref}
\usepackage{multicol}
\usepackage{algorithm}
\usepackage{algorithmic}
\usepackage{svg}
\usepackage{float}
\usepackage{listings}

\definecolor{codebackground}{rgb}{0.97, 0.97, 0.97}

\newtheorem{definition}{Definition}

\def\ie{\emph{i.e.}, }

\begin{document}

\newcommand{\CliffordT}{Clifford + T }

\title{T-Count Optimizing Genetic Algorithm for Quantum State Preparation}
 \author{
     \IEEEauthorblockN{Andrew Wright\IEEEauthorrefmark{1}, Marco Lewis\IEEEauthorrefmark{1}, Paolo Zuliani\IEEEauthorrefmark{2}, Sadegh Soudjani\IEEEauthorrefmark{3}\\}
     \IEEEauthorblockA{\IEEEauthorrefmark{1}Newcastle University, Newcastle upon Tyne, UK\\}
     \IEEEauthorblockA{\IEEEauthorrefmark{2}Università degli Studi di Roma ``La Sapienza", Rome, Italy\\}
     \IEEEauthorblockA{\IEEEauthorrefmark{3}Max Planck Institute for Software Systems, Kaiserslautern, Germany}
 }
\maketitle

\begin{abstract}
    Quantum state preparation is a crucial process within numerous quantum algorithms, and the need for efficient initialization of quantum registers is ever increasing as demand for useful quantum computing grows. The problem arises as the number of qubits to be initialized grows, the circuits required to implement the desired state also exponentially increase in size leading to loss of fidelity to noise. This is mainly due to the susceptibility to environmental effects of the non-Clifford T gate, whose use should thus be reduced as much as possible. In this paper, we present and utilize a genetic algorithm for state preparation circuits consisting of gates from the \CliffordT gate set and optimize them in T-Count as to reduce the impact of noise. Whilst the method presented here does not always produce the most accurate circuits in terms of fidelity, it can generate high-fidelity, non-trivial quantum states such as quantum Fourier transform states. In addition, our algorithm does automatically generate fault tolerantly implementable solutions where the number of the most error prone components is reduced. We present an evaluation of the algorithm when trialed against preparing random, Poisson probability distribution, W, GHZ, and quantum Fourier transform states. We also experimentally demonstrate the scalability issues as qubit count increases, which highlights the need for further optimization of the search process.
\end{abstract}
\begin{IEEEkeywords}
Genetic Algorithms, Quantum Computing, State Preparation, T-Count, \CliffordT
\end{IEEEkeywords}

\section{Introduction}
    The state preparation problem boils down to finding a quantum circuit that efficiently prepares a given target quantum state starting from the all-zero state, \ket{0...0}.
    State preparation plays a major role in many of the algorithms that promise to provide computational advantages over their classical versions. 
    One of the foremost areas where state preparation is required is within the Quantum Fourier Transform (QFT)~\cite{coppersmith2002approximate}.
    The QFT itself is utilized in important quantum algorithms such as Quantum Phase Estimation (QPE)~\cite{kitaev1995quantum} and then by extension, algorithms aiming to provide solutions to the integer factoring problem~\cite{shorPolynomialTimeAlgorithmsPrime1999}.

    State preparation involves initializing the quantum system in such a way that information can be represented at once through superposition and entanglement.
    Therefore, the state preparation problem and the efficiency of its solutions are a significant area of interest, especially within the current era of Noisy Intermediate-Scale Quantum (NISQ) \cite{Preskill-2018} computers where noise is prevalent.

    There have been various methods employed for solving the state preparation problem, such as methods based on the Schmidt decomposition and initial disentanglement of qubits \cite{araujoLowrankQuantumState2023}, techniques utilizing ancillary qubits for storing entanglement information and reductions in computation time in exchange of space \cite{araujoDivideandconquerAlgorithmQuantum2021}, and efficient methods for the preparation of specific quantum states \cite{gleinigEfficientAlgorithmSparse2021}.

    Additionally, there have been many quantum circuit design/optimization techniques based upon evolutionary computation methods \cite{spectorFindingBetterthanclassicalQuantum1999,spectorGeneticProgrammingQuantum1998,sunkelGA4QCOGeneticAlgorithm2023}; and two genetic based methods for quantum state preparation \cite{creeveyGASPGeneticAlgorithm2023,rindellExploringOptimalityApproximate2023}.
    The work \cite{rindellExploringOptimalityApproximate2023} iteratively optimizes state preparation circuits using a genetic algorithm to explore the advantages gained from using approximate state methods on NISQ era machines over exact state preparation techniques when preparing Haar random states. At the same time, \cite{creeveyGASPGeneticAlgorithm2023} presents a more general genetic approach for producing state preparation circuits and their evaluation against the Shende, Bullock, Markov (SBM) method \cite{shendeSynthesisQuantumLogic2005} experimentally implemented by IBM \cite{SummaryQuantumOperations}. Finally, a recent work \cite{Synthetiq} presents a simulated annealing-based approach for the general quantum circuit synthesis problem, of which state preparation can be seen as a special case.
    
    One of the main issues with quantum state preparation algorithms is that the efficiency is based upon the type of quantum state being prepared. This is the reason why many state preparation algorithms are presented for specific quantum states such as Haar random states or probability distribution representations. To this extent, we present an evaluation of the effectiveness of evolutionary techniques in approximate state preparation for a variety of quantum state types as a consequence of evaluating the \CliffordT gate set. 

    Overall, we find varied results for how well the genetic algorithm performs when trialed against preparing different state types. However, as expected, when input size increases, it becomes much harder for the genetic algorithm to find high fidelity solutions, with the exception of the QFT state, which maintains a high fidelity for high qubit counts. For all states however, the genetic algorithm is able to find solution circuits with low T-Count and gate count that result in a high fidelity and additionally, outperform (in the context of circuit depth) the SBM method after transpilation for IBM Quantum systems. 
    
    Our paper is organized as follows: Section~\ref{sec:background} formally defines the state preparation problem and provides some background on evolutionary computation and discusses the reasons for T-Count optimization. Section~\ref{sec:algorithm} describes the genetic algorithm implementation in detail before Section~\ref{sec:results} presents the results obtained by the algorithm, and the paper ends with the conclusions and future work in Section~\ref{sec:conclusion}.

\section{Background}
\label{sec:background}
    \subsection{Quantum State Preparation and Fault Tolerance} \label{QSP}
        Classically, loading data into a circuit is a trivial operation, within quantum computers however, this problem is made much more difficult when mixed states of super-positions or entangled qubits are involved. In the current NISQ era, with a limited number of operations and time to be able to usefully load data into the machine, the efficiency of how this is performed is critical to ensure utilizing the advantage provided by quantum computers.
        
        Formally, the problem of quantum state preparation is defined as follows. 
        \begin{definition}[State Preparation Problem]\label{theproblem}
        Given some finite universal gate set $U$ and a {\em target} $n$-qubit state $\ket{\phi}$, find a quantum circuit $C$ comprised of gates only from $U$ that maps the zero state $\ket{0^n}$ to the target state $\ket{\phi}$:
        \begin{equation*}
        \label{Formal definition}
        C\ket{0^n} \to \ket{\phi}.
        \end{equation*}
        \end{definition} 

        Typically, the utilized gate set corresponds to the native gate set of the real-world quantum computer that the solution is destined to be executed on. Additionally, most of these gate sets are labelled as universal gate sets, in which they can be used to reproduce the effect of any unitary transformation either exactly, or approximately. One of the most common universal gate sets for approximately reproducing transformation effects is the \CliffordT set (Fig. \ref{fig:Clifford+T}).

        \begin{figure}
        \centering
            \begin{subfigure}{0.45\textwidth}
                \[
                    \frac{1}{\sqrt{2}}
                    \begin{bmatrix}
                    1 & 1 \\
                    1 & -1
                    \end{bmatrix}
                    \raisebox{\height}{\hspace{10px}$=$}
                    \raisebox{24px}{\hspace{10px}
                    \Qcircuit @C=1em @R=0.8em @!R { \\
                    & \gate{\mathrm{H}} & \qw
                    }}
                \]
                \subcaption{Hadamard gate}
            \end{subfigure}
            \begin{subfigure}{0.45\textwidth}
                \[
                    \begin{bmatrix}
                        1 & 0 \\
                        0 & i
                    \end{bmatrix}
                    \raisebox{\height}{\hspace{10px}$=$}
                    \raisebox{24px}{\hspace{10px}
                    \Qcircuit @C=1em @R=0.8em @!R { \\
                    & \gate{\mathrm{S}} & \qw
                }}
                \]
                \subcaption{Phase/S gate}
            \end{subfigure}
            \\ \vspace{20px}
            \begin{subfigure}{0.45\textwidth}
                \[
                \begin{bmatrix}
                    1 & 0 & 0 & 0 \\
                    0 & 1 & 0 & 0 \\
                    0 & 0 & 0 & 1 \\
                    0 & 0 & 1 & 0
                \end{bmatrix}
                \raisebox{\height}{\hspace{10px}$=$}
                \raisebox{24px}{\hspace{10px}
                \Qcircuit @C=1em @R=0.8em @!R { \\
                & \ctrl{1} & \qw \\
                & \targ & \qw \\
                }}
                \]
                \subcaption{CNOT gate (with the control qubit being above the target)}
                \label{fig1: CNOT}
            \end{subfigure}
            \begin{subfigure}{0.45\textwidth}
                \[
                \begin{bmatrix}
                    1 & 0 \\
                    0 & e^{i\pi/4}
                \end{bmatrix}
                \raisebox{\height}{\hspace{10px}$=$}
                \raisebox{24px}{\hspace{10px}
                \Qcircuit @C=1em @R=0.8em @!R { \\
                & \gate{\mathrm{T}} & \qw
                }}
                \]
                \subcaption{T gate}
            \end{subfigure}
            \caption{The \CliffordT gate set and the corresponding circuit gate representations.}
            \label{fig:Clifford+T}
        \end{figure}

        Due to the fact that the current quantum computing technologies require extremely precise environmental conditions to operate effectively, it may be required that error correcting measures are utilized in order to reduce the loss of fidelity to noise. Many of the proposed error correction methods require the use of fault tolerant quantum gates which is something the \CliffordT gate set can provide over other universal gate sets. This is what makes the \CliffordT set very desirable for large computations.  
        For a more in-depth introduction into the different error correction/fault tolerance methods, we refer the reader to \cite[Chapter 10]{nielsenQuantumComputationQuantum2010}. 

    \subsection{Evolutionary Inspired Algorithms} \label{EIA}
        With a basis in the theory of natural evolution, evolutionary optimization algorithms aim to iteratively improve solutions to hard computational problems. In the context of this work, a subdomain of evolutionary inspired algorithms, genetic programming, is utilized to solve (approximately) the state preparation problem.

        Traditionally, there are various different genetic methods that are performed during the algorithm, but a general outline of the sequence of a genetic algorithm is as follows.
        Initially, a set of individuals is randomly generated and their respective fitness as solutions is calculated. Then, mutation and crossover operations are performed, and the generated offspring is again evaluated for fitness. These mutation and crossover methods modify solutions to change their structure to increase the chance of producing better solutions. Generally, the algorithm will produce more offspring than were present in the previous generation. This allows for an increase in genetic diversity and can result in a larger percentage of the individuals being more suitable as a solution. From these generated individuals, pre-defined selection methods are performed to reduce the population back down to its original size and therefore potentially removing undesirable solutions. The algorithm then repeats from the mutation/crossover stage until a desired fitness is achieved or the maximum number of rounds is reached, whichever occurs first. 
        For further information on genetic algorithms and the area of evolutionary computation, we refer the reader to \cite{Eiben2015} and \cite{yuIntroductionEvolutionaryAlgorithms2010}.

\subsection{Previous Works}
        Quantum circuit synthesis has been a major topic of research since as early as 1995 \cite{knillApproximationQuantumCircuits1995} with the process of efficient state preparation being a main focus \cite{shendeQuantumCircuitsIncompletely2005}. Furthermore, the use of evolutionary computation techniques has also widely been explored for the synthesis and implementation of specific quantum unitary operations and has achieved promising results~\cite{hutsellApplyingEvolutionaryTechniques2007,spectorFindingBetterthanclassicalQuantum1999,spectorGeneticProgrammingQuantum1998}.
        
        The use of genetic techniques in the preparation of quantum computers has largely gone unfulfilled, only being analyzed by two applications recently \cite{creeveyGASPGeneticAlgorithm2023,rindellExploringOptimalityApproximate2023}. This may be due to the computationally expensive nature of evolutionary and quantum simulation methods, which realistically restrict the advantage they potentially have over other rules-based methods. Both of these techniques utilize evolutionary approaches for the development of state preparation circuits. Rindell et al. \cite{rindellExploringOptimalityApproximate2023} iteratively optimize circuits generated by the LRSP algorithm presented in \cite{araujoLowrankQuantumState2023}, in contrast to the more ``from-the-ground-up" approach presented by Creevey et al. \cite{creeveyGASPGeneticAlgorithm2023}. As \cite{creeveyGASPGeneticAlgorithm2023} and \cite{rindellExploringOptimalityApproximate2023} are both using alternative gate sets to \CliffordT they do not optimize for T-Count as we have done. Instead, \cite{rindellExploringOptimalityApproximate2023} focuses more on circuits accounting for native gate sets and the qubit connectivity of a specific quantum computer in question. In \cite{creeveyGASPGeneticAlgorithm2023}, as the basis gate set involves the three axis rotation gates and CNOT, optimization within the algorithm is performed to find the optimal angle, $\theta$, for each gate. We aim to expand into the optimization area and provide useful tools to be able to automatically synthesize circuits that perform a similar task whilst evaluating the effectiveness of these evolutionary methods in different state preparation tasks.
\subsection{Optimizing for T-Count}
\label{subsec:OpTCount}
    As mentioned in Section \ref{QSP}, the universal \CliffordT gates all have fault tolerant/error identifying implementations. This capacity makes the gate set extremely important in the context of state preparation as quantum computers can then be placed into the required states without propagating errors throughout the following operations, drastically increasing the usability of NISQ-era quantum computers. 
    
    Unfortunately, one current problem with fault tolerance and error correction implementations is their requirement for logical (\ie error-free) qubits. Dependent upon the class of fault tolerant method implemented, the number of physical qubits per logical qubit varies, but the number of required physical qubits increases substantially compared to non-fault tolerant/error correcting methods.
            
    A major problem with the \CliffordT implementation is the susceptibility to noise and the subsequent error rate of the T gate. Ref.~\cite{harper2017estimating} utilizes randomized benchmarking (RB) to estimate the fidelity of Clifford group gates and then subsequently the fidelity of T gates when considering different types of error. On average, for Clifford gates fidelity was found to be $99.67\%$ whereas T gate fidelities were estimated to be $98.6\%$ but were found to be closer to $98.7\%$. For fault-tolerant T gate implementations, the error rate is more pronounced due to the additional space-time overhead needed for the methods used to build the non-Clifford gates. Research into different methods for building these non-Clifford gates such as the T gate has shown reduced error rates during the subsequent computation \cite{Piveteau_2021}.

    In addition to these methods, a large amount of time and effort have been devoted to researching strategies to simply reduce the total number of T gates (T-Count) and aiming to apply these gates at the same time as much as possible (T-Depth) \cite{amyTCountOptimizationReed2019,debeaudrapFastEffectiveTechniques2020,kissingerReducingTcountZXcalculus2020,thapliyalQuantumCircuitDesigns2021}. Therefore, within the work presented here, we aim to genetically produce state preparation circuits that utilize the \CliffordT gate set whilst aiming to reduce the total number of T gates and overall length of the quantum circuit.

\section{The Algorithm}
\label{sec:algorithm}
    Within this section, we outline the implementation of our genetic algorithm and its constituent parts. For reference, the full code repository for our \textbf{T}-count \textbf{O}ptimizing \textbf{G}enetic \textbf{A}lgorithm for quantum \textbf{S}tate preparation (TOGAS) is available on Zenodo \cite{TOGASZenodoRepo}.
    As mentioned in Section \ref{EIA}, we utilize a genetic algorithm to produce quantum circuits that aim to result in some desired state $\ket{\phi}$. 
    We will discuss the different quantum states and their arrangements that our genetic algorithm was trialed against within Section~\ref{sec:results}.
    
    Fig.~\ref{fig:GAPseudocode} shows the pseudocode of the implemented genetic algorithm.
    Within the algorithm, initially $2\times$\emph{pop\_size} individuals are generated with random solution circuits (\emph{pop\_size} is the desired population size). Each individual is evaluated for fitness for use within the subsequent selection method followed by the generational simulation beginning on line 6. Each generation of the algorithm is represented by one execution of the for loop. Initially, the population list is duplicated to produce double the amount of individuals. The while loop on line 9 is used to mate/crossover individuals. Similarly, the for loop on line 15 mutates the individuals. These new individuals are then reevaluated for fitness and the population is then reduced back down to \emph{pop\_size} individuals in preparation for the next generation.

\def\offspring{\textit{offspring}}
\begin{figure}[t]
    \begin{algorithmic}[1]
    \REQUIRE $no\_gens$ (number of generations), $pop\_size$ (population size), $mutpb$ (probability of mutation), $cxpb$ (probability of crossover)
        \STATE $population \gets generate\_individuals(2*pop\_size)$
        \FOR{$i \gets 0$ to $length(population)$}
            \STATE $population[i].fitness \gets evaluation(population[i])$
        \ENDFOR
        \STATE $\offspring \gets selection(population, pop\_size)$
        \FOR{$generation \gets 0$ to $no\_gens$}
            \STATE $i \gets 1$
            \STATE $\offspring \gets \offspring * 2$
            \WHILE{$i < length(\offspring)$}
                \IF{$random\_value() < cxpb$}
                    \STATE $\offspring[i-1], \offspring[i] \gets mate(\offspring[i-1], \offspring[i])$
                \ENDIF
                \STATE $i \gets i + 2$
            \ENDWHILE
            \FOR{$i \gets 0$ to $length(\offspring$)}
                \IF{$random\_value() < mutpb$}
                    \STATE $\offspring[i] \gets mutate(\offspring[i])$
                \ENDIF
            \ENDFOR
            \FOR{$i \gets 0$ to $length(\offspring)$}
                \STATE $\offspring[i].fitness \gets evaluation(\offspring[i])$
            \ENDFOR
            \STATE $\offspring \gets selection(\offspring, pop\_size)$
        \ENDFOR
    \end{algorithmic}
    \caption{Genetic Algorithm Pseudocode}
    \label{fig:GAPseudocode}
\end{figure}
        
    \subsection{Implementation Details}
        The genetic algorithm was developed and implemented with Python and the evolutionary computation library DEAP \cite{DEAPDocumentationDEAP}. Additionally, the quantum simulation and subsequent fitness evaluation was performed using the Qiskit library \cite{Qiskit} provided by IBM. 
        
        As the genetic algorithm was designed to optimize for T-Count and circuit complexity, the resultant genetic algorithm is represented as a multi-objective algorithm that gives each individual a fitness rating corresponding to the different optimization classes.

        \subsection{Solution Representation}\label{sec:solution-rep}
            When discussing genetic algorithms for circuit generation, it is often the case that the circuit is decomposed down into a sequence of its corresponding operations. This allows for easier access to the operations of the circuit to apply mutation or crossover steps within their correlated methods. Within TOGAS, this is still the case. Single qubit gates from Fig.~\ref{fig:Clifford+T} are represented in a manner similar to the following,
            \[
                \textit{Hadamard} = [\textit{`HGate'}, [\textit{Target\_Qubit}]],
            \]
            \[
                \textit{Phase(S)} = [\textit{`SGate'}, [\textit{Target\_Qubit}]],
            \]
            \[
                T(\pi/8) = [\textit{`TGate'}, [\textit{Target\_Qubit}]].
            \]
            With the two-qubit controlled-not gate represented as
            \[
                \textit{CNOT} = [\textit{`CNOT'}, [\textit{Control\_Qubit}, \textit{Target\_Qubit}]].
            \]
            Circuits are then sequences of these representations.

        \subsection{Multi-Objective Fitness}
            Due to the nature of state preparation and solution optimization, it is suitable to define the fitness of a circuit solution as a combination of its performance against multiples objectives, where each objective is weighted to reduce/increase its overall effect on the final performance metric. Within DEAP however, the multiple objectives are kept separate, and solutions are evaluated individually. This multi-objective setup has a larger effect within the different selection methods and as such will be discussed more in Section~\ref{subsec:Discussion}. (For an in-depth introduction to multi-objective genetic algorithms see \cite{Eiben2015,yuIntroductionEvolutionaryAlgorithms2010}.)

            One thing that should be made explicitly clear is how solutions are compared against the desired state. The closeness of two quantum states can be measured in two major ways: state fidelity and trace distance. For state vector comparison performed here, we utilize fidelity as is common within evolutionary algorithms for quantum computing.
            (For more in-depth information on fidelity we refer the reader to \cite[Chapter 9]{nielsenQuantumComputationQuantum2010}.)
            When comparing statevectors rather than density matrices, the fidelity between two density matrices $\rho_1, \rho_2$ is
            \begin{equation*}
                F(\rho_1, \rho_2) = Tr[\sqrt{\sqrt{\rho_{1}}\rho_2\sqrt{\rho_1}}]^2,
                \label{EQ: State Fidelity}
            \end{equation*}
            which can be simplified to
            \begin{equation*}
                F(\ket{\phi}, C\ket{0^n}) = |\bra{\phi}C\ket{0^n}|^2,
                \label{EQ: Simplified State Fidelity}
            \end{equation*}
            where $\ket{\phi}$ is the target state to be produced by the quantum state preparation circuit $C$.
            Within TOGAS, fidelity is calculated as a comparison with the generated noiseless state against the expected state (i.e., in an ideal environment).
            
        \subsection{Crossover}
            In the following, we use the terms individual and (quantum) circuit interchangeably.
            In addition to individual mutation, crossover is one of the main processes that produces the behavior of a genetic algorithm. A significant portion of algorithm performance comes in part from the genetic diversity imparted from individual crossover. It is worth noting that the crossover procedure is also occasionally known as ``reproduction" in that two solutions (circuits) are selected, and their traits are combined in such a way as to produce a set number of offspring. Moreover, at genetic algorithm initialization, a probability parameter is provided to restrict the number of crossover operations. This aims to reduce redundant costly solution evaluations and prevents a situation in which healthy solutions are lost and average solution fitness decreases.

            There are numerous crossover methods that offer different approaches in combining two individuals. Within the tool presented here, we provide the ability to utilize one of the following methods that are provided by DEAP: One Point, Two Point, Messy One Point, and Uniform crossover \cite{DEAPDocumentationDEAP}. 

            \subsubsection{One Point Crossover} 
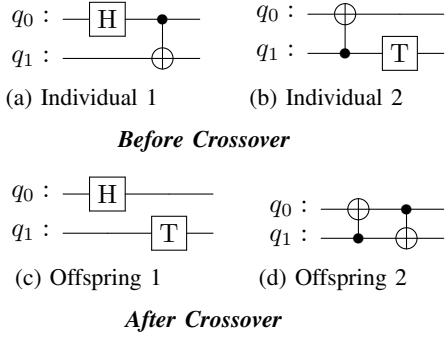
\begin{figure}[t]
                \begin{subfigure}{\linewidth}
                    \centering
                    \subcaptionbox{Individual 1\label{Before Ind 1}}{
                    \scalebox{1}{
                    \Qcircuit @C=1.0em @R=0.2em @!R { \\
                            \nghost{{q} :  } & \lstick{{q_0} :  } & \gate{\mathrm{H}} & \ctrl{1} & \qw\\ 
                            \nghost{{q} :  } & \lstick{{q_1} :  } & \qw & \targ & \qw\\}}}
                    \subcaptionbox{Individual 2\label{Before Ind 2}}{
                    \scalebox{1}{
                    \Qcircuit @C=1.0em @R=0.2em @!R { \\
                            \nghost{{q} :  } & \lstick{{q_0} :  } & \targ & \qw & \qw\\ 
                            \nghost{{q} :  } & \lstick{{q_1} :  } & \ctrl{-1} & \gate{\mathrm{T}} & \qw\\}}}
                    \subcaption*{\textbf{\textit{Before Crossover}}}
                \end{subfigure}
                \begin{subfigure}{\linewidth}
                    \centering
                    \subcaptionbox{Offspring 1\label{After Ind 1}}{
                    \scalebox{1}{
                    \Qcircuit @C=1.0em @R=0.2em @!R { \\
                            \nghost{{q} :  } & \lstick{{q_0} :  } & \gate{\mathrm{H}} & \qw & \qw\\ 
                            \nghost{{q} :  } & \lstick{{q_1} :  } & \qw & \gate{\mathrm{T}} & \qw\\}}}
                    \subcaptionbox{Offspring 2\label{After Ind 2}}{
                    \scalebox{1}{
                    \Qcircuit @C=1.0em @R=0.2em @!R { \\
                            \nghost{{q} :  } & \lstick{{q_0} :  } & \targ & \ctrl{1} & \qw\\ 
                            \nghost{{q} :  } & \lstick{{q_1} :  } & \ctrl{-1} & \targ & \qw\\}}}
                    \subcaption*{\textbf{\textit{After Crossover}}}
                \end{subfigure}
                \caption{Example of One Point Crossover for two circuits}
                \label{fig:Crossover Example}
            \end{figure}

Typically the most often crossover method used due to its simplicity and effectiveness. Traditionally within this method, the first half of the first circuit is combined with the second half of the second circuit to produce the first child circuit. The second child will then comprise of the second half of the first individual and the first half of the second individual {\em etc}. Specifically, the crossover point is given to be the center of the individual's circuit (in the array representation of Section \ref{sec:solution-rep}), that is: $\text{length}(individual)/2$, where length() returns the total number of gates of the circuit. 
            An example is shown in Fig. \ref{fig:Crossover Example}.
            
            \subsubsection{Two Point Crossover} Two point crossover works in a similar way to one point in that, two points are randomly chosen within the individual and the gate operations contained with those two points are swapped to produce two new individuals. 
        
            \subsubsection{Messy One Point Crossover} In the same way as traditional one point crossover, the produced offspring consist of the combined genes (quantum circuit portions) from the parent individuals. The difference, however, is the point of crossover, which for messy one point is a random point from either individual. This has the effect of producing offspring of differing lengths to the parents, which can be more beneficial for finding solutions within the context of circuits and as such we utilize this method for TOGAS evaluation.

            \subsubsection{Uniform Crossover} Uniform crossover works in a much dissimilar way to the previously described procedures. Circuits are iterated over simultaneously, and isolated gates are swapped with some probability. This produces two individuals with an equivalent gate count to the parents.

        \subsection{Mutation}
            Once more, mutation of a circuit, similar to crossover, is a major driving factor of change within solutions during the execution of the genetic algorithm. There are many different mutation operations, and they are all dependent upon the type of problem the genetic algorithm is aiming to solve. In our context, the different types of mutation are fairly unambiguous due to the limited number of changes that can be applied to a circuit.

\begin{figure}
                \centering
                \subcaptionbox{Before mutation\label{Before Mutation}}{
                \scalebox{1}{
                \Qcircuit @C=1.0em @R=0.2em @!R { \\
                        \nghost{{q} :  } & \lstick{{q_0} :  } & \gate{\mathrm{H}} & \qw & \ctrl{1} & \qw\\ 
                        \nghost{{q} :  } & \lstick{{q_1} :  } & \qw & \qw & \targ & \qw\\}}}
                \subcaptionbox{After mutation\label{After Mutation}}{                
                \scalebox{1}{
                \Qcircuit @C=1.0em @R=0.2em @!R { \\
                        \nghost{{q} :  } & \lstick{{q_0} :  } & \gate{\mathrm{H}} & \qw & \ctrl{1} & \qw & \qw & \qw\\ 
                        \nghost{{q} :  } & \lstick{{q_1} :  } & \qw & \qw & \targ & \qw & \gate{\mathrm{T}} & \qw\\}}}
                \caption{Example of circuit mutation with T gate addition}
                \label{fig:Mutation Example}
            \end{figure}
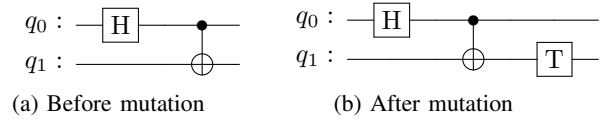

            Additionally, within the genetic algorithm setup and similarly to crossover, mutation probability is specified to prevent the algorithm devolving down into a random solution search.
            We now describe the different mutation operations that can be performed within TOGAS and the potential effects they have for solution performance:
            \if11
            \begin{itemize}
                \item \textbf{Gate Positioning:} Randomly select a gate within the individual and change the target/control qubit to a different qubit in the system.
                \item \textbf{Gate Addition:} Randomly, at some position within the individual, insert a new randomly chosen gate from the gate set utilized.
                \item \textbf{Gate Deletion:} Remove a gate from the individual at some randomly chosen position.
                \item \textbf{Switching:} Replace a random gate from the sequence with a different randomly chosen gate.
                \item \textbf{Sequence Insertion:} Randomly generate a sequence of no more than 25 gates and insert it into a random position within the individual.
                \item \textbf{Sequence Deletion:} Remove up to 25 gates between some randomly chosen position and the end of the individual.
                \item \textbf{Circuit Optimization:} This is the costliest mutation. The entire circuit is iterated over and all gates that result in the same effect as an identity operation are removed. For example, see the two subsequent Hadamard gates in Fig.~\ref{fig:Hadamard Identity}.
                
                Circuit optimization vastly reduces both the depth and overall gate count of generated solutions. It also aids in reducing the average solution gate length which can increase evaluation speed and consequently algorithm efficiency. Additionally, as the implemented gate set does not include the inverse gates, it is possible for sequential applications of rotation gates to be used where a single inverse gate should be utilized instead. (For example, 7 sequential T gates are exactly equivalent to one T$^\dag$ gate.)
            \end{itemize}
            One of these mutation operations is randomly chosen and subsequently applied to the individual. In Fig.~\ref{fig:Mutation Example} we give a simple example of circuit mutation.
            \else
            \vspace{10px}
            \newline
            \textbf{Gate Positioning:} Randomly select a gate within the individual and change the target/control qubit to a different qubit in the system.
            \vspace{10px}
            \newline
            \textbf{Gate Addition:} Randomly, at some position within the individual, insert a new randomly chosen gate from the gate set utilized.
            \vspace{10px}
            \newline
            \textbf{Gate Deletion:} Remove a gate from the individual at some randomly chosen position.
            \vspace{10px}
            \newline
            \textbf{Switching:} Replace a random gate from the sequence with a different randomly chosen gate.
            \vspace{10px}
            \newline
            \textbf{Sequence Insertion:} Randomly generate a sequence of no more than 25 gates and insert it into a random position within the individual.
            \vspace{10px}
            \newline
            \textbf{Sequence Deletion:} Remove $x \leq 25$ gates between some randomly chosen position and the end.
            \vspace{10px}
            \newline
            \textbf{Circuit Optimization:} This is the most costly mutation. The entire circuit is iterated over and all gates that result in the same effect as an identity operation are removed. For example, see the two subsequent Hadamard gates in Fig. \ref{fig:Hadamard Identity}.
            
            Circuit optimization vastly reduces both the depth and overall gate count of generated solutions. It also aids in reducing the average solution gate length which can increase evaluation speed and consequently algorithm efficiency. Additionally, as the implemented gate set does not include the inverse gates, it is possible for sequential applications of rotation gates to be used where a single inverse gate should be utilized instead. (For example, 7 sequential T gates are exactly equivalent to one T$^\dag$ gate.)
            One of these mutation operations is randomly chosen and subsequently applied to the individual. In Fig.~\ref{fig:Mutation Example} we give a simple example of circuit mutation.
            \fi
            \begin{figure}
                \centering
                \hspace{-27px}
                \scalebox{0.85}{
                \Qcircuit @C=1.0em @R=0.2em @!R { \\
                        \nghost{{q} :  } & \lstick{{q} :  } & \gate{\mathrm{H}} & \gate{\mathrm{H}} \barrier[0em]{0} & \qw & \meter & \qw & \qw\\
                        \nghost{\mathrm{{meas} :  }} & \lstick{\mathrm{{meas} :  }} & \lstick{/_{_{1}}} \cw & \cw & \cw & \dstick{_{_{\hspace{0.0em}0}}} \cw \ar @{<=} [-1,0] & \cw & \cw\\
                \\ }}
                \raisebox{-6\height}{\hspace{17px}$=$}
                \scalebox{0.85}{
                \Qcircuit @C=1.0em @R=0.2em @!R { \\
                        \nghost{{q} :  } & \lstick{{q} :  } \barrier[0em]{0} & \qw & \meter & \qw & \qw\\
                        \nghost{\mathrm{{meas} :  }} & \lstick{\mathrm{{meas} :  }} & \lstick{/_{_{1}}} \cw & \dstick{_{_{\hspace{0.0em}0}}} \cw \ar @{<=} [-1,0] & \cw & \cw\\
                \\}}
                \caption{Circuit optimization: Hadamard identity}
                \label{fig:Hadamard Identity}
            \end{figure}
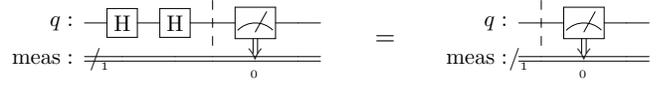

        \subsection{Selection}
            The unambiguous selection process involves performing the action of selecting individuals from the population for crossover and mutation operations. Often, individual selection is based upon fitness. For methods such as best and tournament selection, the fittest individuals are selected for crossover/mutation. In other cases, such as roulette wheel selection, the probabilities of an individual being chosen are proportional to the individuals' fitness. This probability can lead to lower fitness solutions proceeding on to the next generation which can assist in improving genetic diversity.

            Additionally, it is worth to note that due to the multi-objective nature of state preparation, some selection methods do not account for the additional objectives and are only based upon the first objective, which in the context of TOGAS is state fidelity.

            Due to the method of multi-objective fitness declaration utilized within DEAP, individuals' fitness is given as a tuple of values, each corresponding to an algorithm objective. Within TOGAS, individuals are evaluated in three main areas: fidelity (closeness to desired state), gate count of solution (number of gates present within the circuit), and T-Count (number of T gates present within the circuit). As such, the general format of an individual's fitness is given as follows: 
            \begin{equation*}
                \{\text{Fidelity, Gate Count, T Count}\}.
                \label{EQ: TOGAS Fitness Structure}
            \end{equation*}
            This format, when evaluated, initially compares fidelity of two states and upon a tie, moves to comparison of the gate counts of both circuits. If at that point both circuits are equivalent, the final objective to be evaluated is total number of T gates present within the circuit.
            This structure has an underlying effect on how the different selection methods operate and the performance they provide. 

            Within TOGAS, numerous methods of selection are implemented or their DEAP counterparts utilized. These methods include Best, Roulette, Tournament, Double Tournament, Random, Worst, Lexicase, and a bespoke best duplication (selBestDuplication) method. As explained in Section 4.3, the effect of selection within a multi-objective genetic algorithm is determined by how the multi-objective method is implemented. From the methods described above, and within TOGAS, only Best, Worst, Lexicase and Best Duplication selections account for all objectives. Tournament, Double tournament, Random, and Roulette only utilize the first objective (fidelity) during selection. 
            
            For those methods that account for all objectives (excluding Lexicase), DEAP performs tuple comparison to determine the performance of solutions in relation to one another. For a highly circuit dependent quantum state comparison such as fidelity, where the returned value is a real number between 0 and 1, it is often the case where no two solutions are of the same fidelity, this can result in an algorithm that often ignores secondary optimization objectives due to this method of tuple comparison. This is not an uncommon issue among multi-objective algorithms and as such, many different methods have been proposed that aim to solve the problem \cite{DebAFastElitistMultiobjective,HelmuthSolvingUncompromising}.

            One of the most problematic selection methods that we trialed was Lexicase selection presented in \cite{HelmuthSolvingUncompromising}. When evaluating quantum circuit fidelity however, Lexicase selection is unlikely to provide any advantage over other methods due to its workings and devolution down to a random search within the DEAP implementation. Lexicase selection does propose some interesting ideas that could be used to develop other selection methods and as such, the {\em best duplication} method was implemented. Within this method, the population is sorted in order, and the fittest ten solutions are selected and duplicated. Additionally, to improve genetic diversity, ten additional random individuals are also selected and duplicated. The main aim of this was to increase the probability of secondary and tertiary objective evaluation by ensuring that more solutions were evaluated to the same fidelity. This aids in the reduction of both T-Count and circuit length whilst still maintaining some level of genetic diversity such that progress can continue.
\section{Results}
\label{sec:results}
In order to evaluate the effectiveness of the genetic algorithm at being able to produce solutions we carried out experiments with the following quantum states: random states generated via Qiskit, states with probability amplitudes defined from a Poisson distribution, the maximally entangled $W$ state \cite{PhysRevA.62.062314}, the GHZ state \cite{GHZ}, and a quantum Fourier transform state \cite{QFT}.

For all experiments, the population size was set to 150 circuits for 20,000 generations with a crossover probability of 1/2 and a mutation probability of 1/4. The machine utilized for the genetic algorithm and subsequent quantum simulation was an Apple M1 MacBook Pro 8-Core CPU @ 3.2GHz and 8GB RAM, running MacOS Sonoma 14.0.

All state (but Poisson) experiments were executed with the input size ranging from 3 to 12 qubits. While the qubit count is low for these experiments, it can still provide some intuition on how this method should scale as target state grows in size without the obstructive increase in computation time.

\subsection{Random Statevectors}
    To gain an insight into the overall performance of the genetic algorithm, it is required that as many different types of statevector are trialed. It is common to simply describe statevectors in terms of sparsity. That is, statevectors which have complex values of zero for a majority of states are described as sparse \cite{gleinigEfficientAlgorithmSparse2021}. These types of statevector are typically the most often used and, fortunately, can be some of the easiest to prepare. In contrast however, statevectors that contain non-zero values for most states are often the hardest to prepare.

    Random statevectors, for the most part, contain mostly non-zero values for each state within the statevector. This results in statevectors that can be difficult to prepare and is the reason random statevectors play an important role in state preparation evaluation techniques.

    To generate the random statevectors used within the evaluation, we utilize the Qiskit \textbf{random\_statevector} method which samples from the uniform Haar measure, similarly to \cite{rindellExploringOptimalityApproximate2023}.

\subsection{Poisson Statevectors}
    Another common type of state preparation problem relates to the initial preparation of algorithms such as Monte Carlo simulation methods, whereby the quantum computer state is required to be a probability distribution. This makes probability distributions a good gauge on how effective state preparation methods are for real world use cases.

    The amplitudes for each state are given by the Poisson formula,
    \begin{equation*}
        \frac{\lambda^xe^{-\lambda}}{x!}
    \end{equation*}
    whereby, $\lambda$ is given to be the average number of occurrences within the distribution; $\lambda = \frac{2^n}{2}$, for $n$ qubits. The generated amplitudes are then normalized such that the sum of all probabilities is equal to one to produce the following state:   
    \begin{equation*}
        \ket{\phi} = \sum_{x=0}^{2^n-1} \frac{\lambda^xe^{-\lambda}}{x!}\ket{x}.
    \end{equation*}

\subsection{W State Statevectors}
    The third type of state used for algorithm analysis is the entangled $W$ state over $n$ qubits
    \begin{equation*}
        W = \frac{1}{\sqrt{n}}(\ket{100...0} + \ket{010...0} + ... + \ket{00...01}).
    \end{equation*}
    The $W$ state is composed of all possible states where only one qubit is in the $\ket{1}$ state with uniform probability. Much like probability distributions, $W$ states are often used as representatives of logical 1 within fault tolerant methods due to their robustness against loss.

\subsection{GHZ Statevectors}
    Another type of statevector that we use for evaluation is the Greenberger–Horne–Zeilinger (GHZ) state. It is of a similar classification to the $W$ state whereby it can be generalized for $n$ qubits as follows:
    \begin{equation*}
        GHZ = \frac{\ket{00...0} + \ket{11...1}}{\sqrt{2}}.
    \end{equation*}
    In general, it is the equal superposition of the all $0$ and all $1$ state.
    
\subsection{QFT Statevectors}
    The quantum alternative to the Discrete Fourier Transform, the QFT over $N$ qubits is defined as following

    \begin{equation*}
        \ket{\phi} \to \frac{1}{\sqrt{N}}\sum_{x=0}^{N-1}e^{2\pi\\i\phi\\x/N}\ket{x}.
    \end{equation*}

    In reference to its uses for algorithm evaluation within this paper, we aim to produce a circuit which mimics the application of QFT to some input state. As QFT application to various input states can produce a multitude of output states, we simply mimic applying QFT to the all 1 $N$-qubit state, $\ket{1 1 ... 1}$, to introduce some aspect of phase into the desired state. This will allow the genetic algorithm to be evaluated on production of more structured, non-random states with additional phase requirements.

\section{Discussion}
\label{subsec:Discussion}
    \begin{table}[t]
        \centering
        \caption{Performance of TOGAS on Random, Poisson, W, GHZ, and QFT state preparation for input size $n$ ranging from 3-6, and 12 qubits.
        The results report mean ± standard deviation over 10 independent runs of TOGAS.
        {\bf Bold} entries highlight experiments that returned low-fidelity circuits ($\leqslant 0.5)$. {\bf\em Bold italics} entries denote experiments that produced
        a large, outlier solution.}
        \vspace{7px}
        \begin{tabular}{p{0.15\linewidth}|p{0.04\linewidth}|p{0.12\linewidth}|p{0.15\linewidth}|p{0.11\linewidth}|p{0.15\linewidth}}
            \hline
            Experiment & $n$ & T Count & Gate Count & Fidelity & Time($s$)\\
            \hline\hline
            Random & 3 & 93.5 ± 160.91 & 232.8 ± 303.5 & 0.9645 ± 0.024 & 7217.43 ± 2848.23 \\
             & 4 & 23.7 ± 21.48 & 80.2 ± 65.01 & 0.831 ± 0.092 & 4959.75 ± 1265.14 \\
             & 5 & 21.2 ± 19.93 & 73.8 ± 48.95 & 0.6525 ± 0.068 & 5029.73 ± 952.16 \\
             & 6 & 33.1 ± 32.31 & 125.4 ± 110.87 & \textbf{0.4149 ± 0.105} & 6127.52 ± 2693.52 \\
             & \textbf{12} & \textbf{68.33} ± \textbf{124.28} & \textbf{150.44} ± \textbf{220.1} & \textbf{0.0092} ± \textbf{0.003} & \textbf{8308.97} ± \textbf{7417.11} \\
            \hline \hline      
            Poisson & 3 & 28.5 ± 45.15 & 100.6 ± 129.55 & 0.9728 ± 0.018 & 3830.99 ± 1124.23 \\
             & 4 & 7.1 ± 9.13 & 45.2 ± 36.45 & 0.913 ± 0.026 & 3270.25 ± 483.91 \\
             & 5 & 4.5 ± 0.81 & 23.2 ± 8.75 & 0.9555 ± 0.0 & 3302.61 ± 289.6 \\
             & 6 & 5.5 ± 10.69 & 29.2 ± 30.32 & 0.8987 ± 0.027 & 3461.5 ± 794.14 \\
            \hline\hline        
            W & 3 & 7.2 ± 4.07 & 36.1 ± 15.95 & 0.976 ± 0.007 & 3665.4 ± 532.42 \\
             & 4 & 8.9 ± 3.88 & 41.0 ± 9.96 & 0.9483 ± 0.085 & 4012.73 ± 418.12 \\
             & 5 & 9.6 ± 6.45 & 42.9 ± 19.08 & 0.718 ± 0.078 & 4047.7 ± 734.72 \\
             & 6 & 12.6 ± 10.98 & 51.1 ± 25.29 & 0.6303 ± 0.087 & 4376.97 ± 807.12 \\
             & \textbf{12} & \textbf{7.33} ± \textbf{4.78} & \textbf{41.67} ± \textbf{13.62} & \textbf{0.2692} ± \textbf{0.037} & \textbf{4587.5} ± \textbf{1318.34} \\
            \hline\hline
            GHZ & \textbf{\textit{3}} & \textbf{\textit{41.4}} ± \textbf{\textit{115.89}} & \textbf{\textit{300.0}} ± \textbf{\textit{842.34}} & 1.0 ± 0.0 & 2676.9 ± 564.94 \\
            & \textbf{\textit{4}} & \textbf{\textit{202.2}} ± \textbf{\textit{594.94}} & \textbf{\textit{398.4}} ± \textbf{\textit{1131.23}} & 1.0 ± 0.0 & 2778.66 ± 670.85 \\
            & 5 & 3.5 ± 2.01 & 16.0 ± 4.52 & 1.0 ± 0.0 & 2967.41 ± 101.3 \\
            & 6 & 2.2 ± 1.78 & 14.2 ± 5.06 & 0.95 ± 0.15 & 2825.61 ± 130.1 \\
            & \textbf{12} & \textbf{0.78} ± \textbf{0.63} & \textbf{2.33} ± \textbf{0.67} & \textbf{0.5} ± \textbf{0.0} & \textbf{2661.78} ± \textbf{203.57} \\
            \hline\hline
            QFT & 3 & 4.0 ± 2.83 & 17.8 ± 5.71 & 1.0 ± 0.0 & 2954.51 ± 162.84 \\
            & 4 & 7.5 ± 9.65 & 34.4 ± 26.47 & 0.9695 ± 0.015 & 3467.44 ± 535.92 \\
            & 5 & 3.5 ± 3.07 & 23.1 ± 12.41 & 0.9588 ± 0.012 & 3420.19 ± 299.06 \\
            & 6 & 4.0 ± 2.45 & 20.7 ± 3.49 & 0.9504 ± 0.0 & 3476.57 ± 214.82 \\
            & 12 & 2.56 ± 1.77 & 27.78 ± 8.31 & 0.9496 ± 0.0 & 4526.07 ± 453.24 \\
            \hline
        \end{tabular}
        \label{Table: ResultsTable}
    \end{table}
    The results presented in Table \ref{Table: ResultsTable} were gathered from ten independent executions of TOGAS in order to get an accurate representation of its performance.
    
    It is often the case that the speed at which perfect solutions are found is utilized to evaluate the implementation. As state preparation is a difficult problem however, it is more useful to evaluate the performance of the genetic algorithm on the fidelity of the states, and the time required to produce a solution circuit. In terms of TOGAS evaluation, the algorithm had a maximum generation limit of 20,000, as from initial trial executions, the genetic algorithm was able to find solutions of fidelity $\geq 0.90$ quickly within this limit for small input sizes.

    Additionally, since the \CliffordT gate set can only realistically be used for approximations, this therefore limits the ability for the genetic algorithm to find solutions within a reasonable time, with reasonable circuit sizes, with high fidelity. Therefore, we terminate the algorithm if the average size of the circuits in the population exceeds 2,000 gates.

    From the results presented in Table \ref{Table: ResultsTable}, it can be seen that TOGAS is able to produce good solutions for any type of statevector when considering low input sizes. This changes as qubit count increases and it is apparent that it becomes much more difficult for the algorithm to find solutions for preparing random states: the average fidelity of best produced circuits (marked in \textbf{bold} in Table~\ref{Table: ResultsTable}) after 20,000 generations is smaller than $ 0.5$. This decrease is seen across all types of statevector but is not as significant as for random statevector preparation.

    From the results for Poisson statevector generation, it can be seen that while differences in fidelity are $\approx 0.1$, the trials of all input sizes reach an acceptable minimum fidelity of $0.89$ with, on average, small gate/T-counts. One thing of interest to note is the abnormally high fidelity and lowest T and gate counts of the 5 qubit Poisson experiments on average. Moreover, the standard deviation of fidelity for this input size is zero since for all ten TOGAS executions, the resulting fidelity was 0.955. This is most likely due to the form of the resulting state whereby it is much easier for the algorithm to reach a certain distribution compared to other input sizes. Additionally, the time to produce the corresponding circuits remains relatively stable at approximately one hour.

    Within preparation of $W$ states, the average fidelity peaks for an input size of 3 qubits and subsequently decreases. As TOGAS is designed to account for T/gate count, it is likely that the algorithm will settle into a local optimum as any attempts at increasing circuit size to find potential assistant circuit components are lost to the selection method. 

    Interestingly, both GHZ and QFT state preparation experiments do not seem to be as affected by fidelity loss as qubit count increases when compared with other experiments. This is understandable for the GHZ states as the structure of the state is relatively simple and the required circuits have ideal circuit sizes of $n+1$ gates. QFT state preparation, however, is much more intriguing as the state is still structured but also includes added phase complexity which one may imagine would increase the complexity of the overall state. 

    Additionally, it is worth mentioning that for 3 and 4 qubit GHZ states (marked in \textbf{\textit{bold italics}}), TOGAS produced a single outlier solution for each. This is noted as the reason for the large standard deviations seen in Table \ref{Table: ResultsTable}. On average, without these outliers, the T-Count and Gate-count are closer to those presented in the 5 and 6 qubit GHZ experiments. For clarity, the results of the 3 and 4 qubit GHZ experiments without the outlier data can be seen in Table \ref{Table: GHZ Table}.

    Whilst ideally, GHZ state preparation should not involve any T gates, TOGAS produced solutions are seen to contain them. This is not optimal but given enough time and freedom, we believe the genetic algorithm will be able to remove these from solutions.

    \begin{table}[t]
        \centering
        \caption{Performance of TOGAS on GHZ state preparation for 3 and 4 qubits without outlier data (9 valid experiments each).
        The results report mean ± standard deviation over 9 independent runs of TOGAS. }
        \vspace{7px}
        \begin{tabular}{p{0.15\linewidth}|p{0.04\linewidth}|p{0.12\linewidth}|p{0.15\linewidth}|p{0.11\linewidth}|p{0.15\linewidth}}
            \hline
            Experiment & $n$ & T Count & Gate Count & Fidelity & Time($s$)\\
            \hline\hline
            GHZ & 3 & 2.78 ± 2.2 & 19.22 ± 4.57 & 1.0 ± 0.0 & 2857.98 ± 163.49 \\
            & 4 & 3.89 ± 3.31 & 21.33 ± 8.56 & 1.0 ± 0.0 & 2985.36 ± 269.84 \\
            \hline
        \end{tabular}
        \label{Table: GHZ Table}
    \end{table}

    Across all experiments, as was expected, the average fidelity decreases (even if only by small amounts) as qubit count increases due to the increase in state complexity. 
    Consequently, this implies that the genetic algorithm would require much more modification for each state attempting to be prepared. In addition, the required changes would most likely vary greatly between states and input sizes which practically prevents the use of evolutionary techniques for the straight synthesis of state preparation circuits. Ideally, this would not be the case for a generic state preparation method. It is therefore likely to be much more practical to use genetic techniques for the development of smaller circuit components much like \cite{spectorGeneticProgrammingQuantum1998,spectorFindingBetterthanclassicalQuantum1999}, or for optimization methods as presented in \cite{rindellExploringOptimalityApproximate2023}.

    Moreover, the time required for genetic methods reduces its usefulness further as for even small input sizes, adequate fidelity circuit generation requires significantly more time when compared to other methods such as SBM \cite{shendeSynthesisQuantumLogic2005}.

    Furthermore, the standard deviation from the mean fidelity highlights the inherent randomness of genetic methods as the search space increases. It is also interesting to note that the time required for the genetic algorithm to finish remains stable with only those experiments that have large averages, such as Random state preparation, differing greatly in time to reach the 20,000 generation limit.

    Finally, we also trialed TOGAS on an input size of 12 qubits for all but the Poisson probability distribution problem due to the inability to generate the full state. As expected, the fidelity drops quickly whilst the overall T/Gate count grows large with the exception of the QFT statevectors, which remains stable and only slightly reduced in fidelity. The time required for the algorithm to complete, particularly for Random statevector preparation, is hampered by the simulation time of the circuit as the average number of gates within the circuit grows. Additionally, as best solutions are duplicated for the following generation, the average time to simulate all individuals increases drastically if the circuit is large.
    
    \subsubsection*{\textbf{Comparison with Other Methods}}
        Whilst the work presented here is of a very similar topic to the work presented in \cite{rindellExploringOptimalityApproximate2023}, and \cite{creeveyGASPGeneticAlgorithm2023}, a comparison between the results presented here in Table \ref{Table: ResultsTable} would be of no real merit due to the different nuances of each method. Differences such as gate set are of most concern and the effect it would have on solutions. Instead, we present the results as an individual evaluation of genetic methods as a technique of state preparation concerning a multitude of general desired states with the added requirement of reduced T-Count.

    \subsubsection*{\textbf{The Restrictiveness of Desired Properties}}
        What is prevalent in Table \ref{Table: ResultsTable} is that, for all experiments, as input size increases, the T-Count and Gate count remain within similar ranges. Ideally, as input size increases, the algorithm would be required to find solutions that utilize an increased number of gates to improve fidelity. Unfortunately, one of the major impacts of TOGAS is the reduction in probability of the genetic algorithm finding these solutions due to the extra restrictiveness placed upon the algorithm for T/Gate count. This for the most part prevents the algorithm from reaching out of a small search space to find higher fidelity solutions. Again, this may be opposed by more systematic changes to the underlying implementation of the genetic algorithm that would better suit the type of states to be prepared.

    \subsubsection*{\textbf{Transpilation for IBM Quantum Computers}}
\begin{figure*}
        \centering
            \begin{subfigure}{0.4\linewidth}
                \centering
                \subcaptionbox{TOGAS Generated 4 Qubit W State after transpilation. Circuit Depth = $50$, Gate Count = $72$\label{TOGAS WState}}[0.75\linewidth]{
                \includegraphics[width=0.95\paperwidth, angle=90]{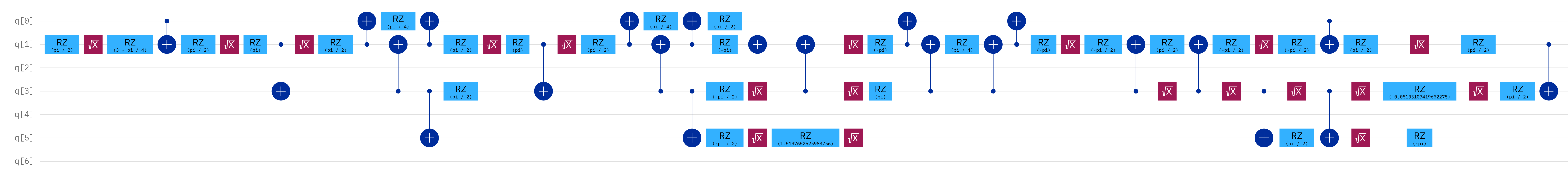}}
            \end{subfigure}
            \begin{subfigure}{0.4\linewidth}
            \centering
                \subcaptionbox{SBM Generated 4 Qubit W State after transpilation. Circuit Depth = $72$, Gate Count = $83$\label{SBM WState}}[0.75\linewidth]{
                \includegraphics[width=0.95\paperwidth, angle=90]{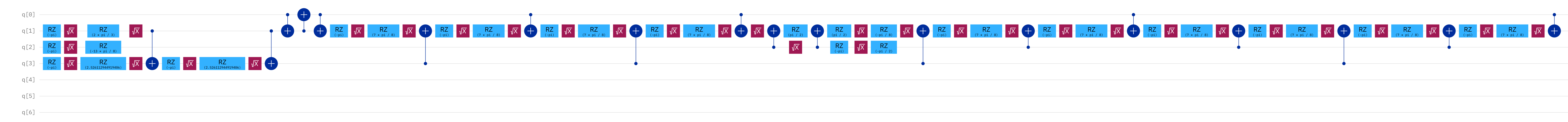}}
            \end{subfigure}
            \caption{Example of TOGAS and SBM~\cite{shendeSynthesisQuantumLogic2005} generated circuits (Fig.~\ref{TOGAS WState} and Fig.\ref{SBM WState} respectively) after transpilation down to the IBM Quantum device native gate set: $\{CX, RZ, \sqrt{X}, X\}$. Genetically generated, \CliffordT circuits shown to have reduced circuit depth after transpilation. Full scale SVG versions of these figures are available in \cite{TOGASZenodoRepo}.}
            \label{fig: Transpiled Circuits}
        \end{figure*}

        We have also verified that the TOGAS-generated circuits with fidelity greater than $0.9$ do produce correct behavior when experimentally implemented on real world quantum devices provided by IBM \cite{IBMQuantumComputing2015}. The circuits generated are transpiled down to sequences of the native quantum operations that the real-world devices support. An example can be seen in Fig. \ref{fig: Transpiled Circuits}, which shows that the transpiled TOGAS circuit (Fig. \ref{TOGAS WState}) is a great deal shorter in terms of depth than those of the default Qiskit SBM method. This may be extremely important to take note of when considering much larger circuits in order to reduce the overall effect of noise. While genetic algorithms may not have a place in full circuit generation, it may be wise to utilize them, much like in \cite{rindellExploringOptimalityApproximate2023} and \cite{sunkelGA4QCOGeneticAlgorithm2023} for circuit optimization instead.

\section{Conclusion}
\label{sec:conclusion}
    We have presented and evaluated a quantum state preparation method based upon evolutionary computation techniques built around the Clifford + T gate set. Our technique aims at producing state preparation circuits that are optimized in terms of output fidelity, T and full gate count. 
    We have evaluated the presented method against multiple different problems as a multi-purpose state preparation method and the presented results indicate that for small input sizes, genetic methods can prove to be useful in generation of state preparation circuits that are as good as or better than non-evolutionary based methods in areas such as circuit depth. Moreover, the results demonstrate the relation of complexity to input size as it scales up. It is also shown that as a multi-purpose state preparation tool, the overall method would need vast improvements in scalability and much like other methods, state preparation of certain state types are shown to be easier than others. Additionally, it may prove more beneficial to utilize evolutionary computation methods for the T-Count optimization of larger circuits generated by other methods such as SBM \cite{shendeSynthesisQuantumLogic2005} and this is something that could be explored. 
 \section*{Acknowledgements}
 M.L.~was supported by the UK Engineering and Physical Sciences Research Council (EPSRC project reference EP/T517914/1). P.Z.~was supported by the project SERICS (PE00000014) under the Italian MUR National Recovery and Resilience Plan funded by the European Union - NextGenerationEU. The research of S. Soudjani was supported by the following grants: EPSRC EP/V043676/1, EIC 101070802, and ERC 101089047.

\bibliographystyle{IEEEtran}
\bibliography{citations}

\begin{thebibliography}{10}
\providecommand{\url}[1]{#1}
\csname url@samestyle\endcsname
\providecommand{\newblock}{\relax}
\providecommand{\bibinfo}[2]{#2}
\providecommand{\BIBentrySTDinterwordspacing}{\spaceskip=0pt\relax}
\providecommand{\BIBentryALTinterwordstretchfactor}{4}
\providecommand{\BIBentryALTinterwordspacing}{\spaceskip=\fontdimen2\font plus
\BIBentryALTinterwordstretchfactor\fontdimen3\font minus
  \fontdimen4\font\relax}
\providecommand{\BIBforeignlanguage}[2]{{%
\expandafter\ifx\csname l@#1\endcsname\relax
\typeout{** WARNING: IEEEtran.bst: No hyphenation pattern has been}%
\typeout{** loaded for the language `#1'. Using the pattern for}%
\typeout{** the default language instead.}%
\else
\language=\csname l@#1\endcsname
\fi
#2}}
\providecommand{\BIBdecl}{\relax}
\BIBdecl

\bibitem{coppersmith2002approximate}
\BIBentryALTinterwordspacing
D.~Coppersmith, ``An approximate {Fourier} transform useful in quantum
  factoring,'' Jan. 2002, arXiv:quant-ph/0201067. [Online]. Available:
  \url{http://arxiv.org/abs/quant-ph/0201067}
\BIBentrySTDinterwordspacing

\bibitem{kitaev1995quantum}
A.~Y. Kitaev, ``Quantum measurements and the abelian stabilizer problem,''
  1995, arXiv:9511026 [quant-ph].

\bibitem{shorPolynomialTimeAlgorithmsPrime1999}
\BIBentryALTinterwordspacing
P.~W. Shor, ``Polynomial-{Time} {Algorithms} for {Prime} {Factorization} and
  {Discrete} {Logarithms} on a {Quantum} {Computer},'' \emph{SIAM Review},
  vol.~41, no.~2, pp. 303--332, 1999, publisher: Society for Industrial and
  Applied Mathematics. [Online]. Available:
  \url{https://www.jstor.org/stable/2653075}
\BIBentrySTDinterwordspacing

\bibitem{Preskill-2018}
\BIBentryALTinterwordspacing
J.~Preskill, ``Quantum {C}omputing in the {NISQ} era and beyond,''
  \emph{{Quantum}}, vol.~2, p.~79, Aug. 2018. [Online]. Available:
  \url{https://doi.org/10.22331/q-2018-08-06-79}
\BIBentrySTDinterwordspacing

\bibitem{araujoLowrankQuantumState2023}
I.~F. Araujo, C.~Blank, I.~C.~S. Araújo, and A.~J. da~Silva, ``Low-rank
  quantum state preparation,'' \emph{IEEE Transactions on Computer-Aided Design
  of Integrated Circuits and Systems}, vol.~43, no.~1, pp. 161--170, 2024.

\bibitem{araujoDivideandconquerAlgorithmQuantum2021}
\BIBentryALTinterwordspacing
I.~F. Araujo, D.~K. Park, F.~Petruccione, and A.~J. da~Silva, ``A
  divide-and-conquer algorithm for quantum state preparation,''
  \emph{Scientific Reports}, vol.~11, no.~1, p. 6329, Mar 2021. [Online].
  Available: \url{https://doi.org/10.1038/s41598-021-85474-1}
\BIBentrySTDinterwordspacing

\bibitem{gleinigEfficientAlgorithmSparse2021}
N.~Gleinig and T.~Hoefler, ``An efficient algorithm for sparse quantum state
  preparation,'' in \emph{2021 58th ACM/IEEE Design Automation Conference
  (DAC)}, 2021, pp. 433--438.

\bibitem{spectorFindingBetterthanclassicalQuantum1999}
L.~Spector, H.~Barnum, H.~Bernstein, and N.~Swamy, ``Finding a
  better-than-classical quantum {AND}/{OR} algorithm using genetic
  programming,'' in \emph{Proceedings of the 1999 {Congress} on {Evolutionary}
  {Computation}-{CEC99} ({Cat}. {No}. {99TH8406})}, vol.~3, Jul. 1999, pp.
  2239--2246.

\bibitem{spectorGeneticProgrammingQuantum1998}
L.~Spector, H.~Barnum, H.~J. Bernstein, and N.~Swamy, ``Genetic programming for
  quantum computers,'' \emph{Genetic Programming}, pp. 365--373, 1998.

\bibitem{sunkelGA4QCOGeneticAlgorithm2023}
L.~Sünkel, D.~Martyniuk, D.~Mattern, J.~Jung, and A.~Paschke, ``{GA4QCO}:
  Genetic algorithm for quantum circuit optimization,'' May 2023,
  arXiv:2302.01303 [quant-ph].

\bibitem{creeveyGASPGeneticAlgorithm2023}
\BIBentryALTinterwordspacing
F.~M. Creevey, C.~D. Hill, and L.~C.~L. Hollenberg, ``{GASP}: a genetic
  algorithm for state preparation on quantum computers,'' \emph{Scientific
  Reports}, vol.~13, no.~1, p. 11956, Jul 2023. [Online]. Available:
  \url{https://doi.org/10.1038/s41598-023-37767-w}
\BIBentrySTDinterwordspacing

\bibitem{rindellExploringOptimalityApproximate2023}
T.~Rindell, B.~Yenilen, N.~Halonen, A.~Pönni, I.~Tittonen, and M.~Raasakka,
  ``Exploring the optimality of approximate state preparation quantum circuits
  with a genetic algorithm,'' \emph{Physics Letters A}, vol. 475, p. 128860,
  2023.

\bibitem{shendeSynthesisQuantumLogic2005}
V.~V. Shende, S.~S. Bullock, and I.~L. Markov, ``Synthesis of quantum logic
  circuits,'' in \emph{Proceedings of the 2005 Asia and South Pacific Design
  Automation Conference}, ser. ASP-DAC '05.\hskip 1em plus 0.5em minus
  0.4em\relax New York, NY, USA: Association for Computing Machinery, 2005, p.
  272–275.

\bibitem{SummaryQuantumOperations}
\BIBentryALTinterwordspacing
``Summary of {Quantum} {Operations} - {Qiskit} 0.44.2 documentation.''
  [Online]. Available:
  \url{https://qiskit.org/documentation/tutorials/circuits/3_summary_of_quantum_operations.html#Arbitrary-initialization}
\BIBentrySTDinterwordspacing

\bibitem{Synthetiq}
\BIBentryALTinterwordspacing
A.~Paradis, J.~Dekoninck, B.~Bichsel, and M.~Vechev, ``Synthetiq: Fast and
  versatile quantum circuit synthesis,'' \emph{Proc. ACM Program. Lang.},
  vol.~8, no. OOPSLA1, 2024. [Online]. Available:
  \url{https://doi.org/10.1145/3649813}
\BIBentrySTDinterwordspacing

\bibitem{nielsenQuantumComputationQuantum2010}
M.~A. Nielsen and I.~L. Chuang, \emph{Quantum Computation and Quantum
  Information: 10th Anniversary Edition}.\hskip 1em plus 0.5em minus
  0.4em\relax Cambridge University Press, 2010.

\bibitem{Eiben2015}
A.~E. Eiben and J.~E. Smith, \emph{Introduction to Evolutionary Computing,
  Second Edition}, ser. Natural Computing Series.\hskip 1em plus 0.5em minus
  0.4em\relax Springer, 2015.

\bibitem{yuIntroductionEvolutionaryAlgorithms2010}
X.~Yu and M.~Gen, \emph{Introduction to Evolutionary Algorithms}, ser. Decision
  Engineering.\hskip 1em plus 0.5em minus 0.4em\relax London: Springer, 2010.

\bibitem{knillApproximationQuantumCircuits1995}
E.~Knill, ``Approximation by quantum circuits,'' 1995, arXiv:9508006
  [quant-ph].

\bibitem{shendeQuantumCircuitsIncompletely2005}
V.~V. Shende and I.~L. Markov, ``Quantum circuits for incompletely specified
  two-qubit operators,'' \emph{Quantum Info. Comput.}, vol.~5, no.~1, p.
  49–57, Jan 2005.

\bibitem{hutsellApplyingEvolutionaryTechniques2007}
S.~R. Hutsell and G.~W. Greenwood, ``Applying evolutionary techniques to
  quantum computing problems,'' in \emph{2007 {IEEE} {Congress} on
  {Evolutionary} {Computation}}, Sep. 2007, pp. 4081--4085, iSSN: 1941-0026.

\bibitem{harper2017estimating}
R.~Harper and S.~T. Flammia, ``Estimating the fidelity of {T} gates using
  standard interleaved randomized benchmarking,'' \emph{Quantum Science and
  Technology}, vol.~2, no.~1, p. 015008, 2017.

\bibitem{Piveteau_2021}
C.~Piveteau, D.~Sutter, S.~Bravyi, J.~M. Gambetta, and K.~Temme, ``Error
  mitigation for universal gates on encoded qubits,'' \emph{Phys. Rev. Lett.},
  vol. 127, p. 200505, Nov 2021.

\bibitem{amyTCountOptimizationReed2019}
M.~Amy and M.~Mosca, ``T-count optimization and {Reed–Muller} codes,''
  \emph{IEEE Transactions on Information Theory}, vol.~65, no.~8, pp.
  4771--4784, Aug 2019.

\bibitem{debeaudrapFastEffectiveTechniques2020}
N.~de~Beaudrap, X.~Bian, and Q.~Wang, ``{Fast and Effective Techniques for
  T-Count Reduction via Spider Nest Identities},'' in \emph{15th Conference on
  the Theory of Quantum Computation, Communication and Cryptography (TQC
  2020)}, ser. Leibniz International Proceedings in Informatics (LIPIcs), S.~T.
  Flammia, Ed., vol. 158.\hskip 1em plus 0.5em minus 0.4em\relax Dagstuhl,
  Germany: Schloss Dagstuhl -- Leibniz-Zentrum f{\"u}r Informatik, 2020, pp.
  11:1--11:23.

\bibitem{kissingerReducingTcountZXcalculus2020}
A.~Kissinger and J.~van~de Wetering, ``Reducing the number of non-{Clifford}
  gates in quantum circuits,'' \emph{Phys. Rev. A}, vol. 102, p. 022406, Aug
  2020.

\bibitem{thapliyalQuantumCircuitDesigns2021}
H.~Thapliyal, T.~S.~S. Varun, E.~Munoz-Coreas, K.~A. Britt, and T.~S. Humble,
  ``Quantum circuit designs of integer division optimizing {T}-count and
  {T}-depth,'' in \emph{2017 IEEE International Symposium on Nanoelectronic and
  Information Systems (iNIS)}, 2017, pp. 123--128.

\bibitem{TOGASZenodoRepo}
\BIBentryALTinterwordspacing
``\BIBforeignlanguage{en}{T-count optimizing genetic algorithm for quantum
  state preparation ({TOGAS})}.'' [Online]. Available:
  \url{https://doi.org/10.5281/zenodo.11444269}
\BIBentrySTDinterwordspacing

\bibitem{DEAPDocumentationDEAP}
\BIBentryALTinterwordspacing
``{DEAP} documentation — {DEAP} 1.4.1 documentation.'' [Online]. Available:
  \url{https://deap.readthedocs.io/en/master/}
\BIBentrySTDinterwordspacing

\bibitem{Qiskit}
A.~Javadi-Abhari, M.~Treinish, K.~Krsulich, C.~J. Wood, J.~Lishman, J.~Gacon,
  S.~Martiel, P.~D. Nation, L.~S. Bishop, A.~W. Cross, B.~R. Johnson, and J.~M.
  Gambetta, ``Quantum computing with {Q}iskit,'' 2024, arXiv:2405.08810
  [quant-ph].

\bibitem{DebAFastElitistMultiobjective}
K.~Deb, A.~Pratap, S.~Agarwal, and T.~Meyarivan, ``A fast and elitist
  multiobjective genetic algorithm: Nsga-ii,'' \emph{IEEE Transactions on
  Evolutionary Computation}, vol.~6, no.~2, pp. 182--197, 2002.

\bibitem{HelmuthSolvingUncompromising}
T.~Helmuth, L.~Spector, and J.~Matheson, ``Solving uncompromising problems with
  lexicase selection,'' \emph{IEEE Transactions on Evolutionary Computation},
  vol.~19, no.~5, pp. 630--643, 2015.

\bibitem{PhysRevA.62.062314}
W.~D\"ur, G.~Vidal, and J.~I. Cirac, ``Three qubits can be entangled in two
  inequivalent ways,'' \emph{Phys. Rev. A}, vol.~62, p. 062314, Nov 2000.

\bibitem{GHZ}
\BIBentryALTinterwordspacing
D.~M. Greenberger, M.~A. Horne, and A.~Zeilinger, \emph{Going Beyond Bell's
  Theorem}.\hskip 1em plus 0.5em minus 0.4em\relax Dordrecht: Springer
  Netherlands, 1989, pp. 69--72. [Online]. Available:
  \url{https://doi.org/10.1007/978-94-017-0849-4_10}
\BIBentrySTDinterwordspacing

\bibitem{QFT}
P.~W. Shor, ``Algorithms for quantum computation: discrete logarithms and
  factoring,'' in \emph{Proceedings 35th Annual Symposium on Foundations of
  Computer Science}, 1994, pp. 124--134.

\bibitem{IBMQuantumComputing2015}
``{IBM} {Quantum} {Computing},'' \url{https://www.ibm.com/quantum}.

\end{thebibliography}
\end{document}